\documentclass[12pt]{article}
\usepackage{
amsmath,
amssymb,
bm,
tikz,
braket,
mathrsfs,
graphicx,
color,
hyperref,
}

\hypersetup{colorlinks,bookmarksopen,bookmarksnumbered,citecolor=black,linkcolor=black,linktocpage,pdfstartview=FitH,urlcolor=black}

\numberwithin{equation}{section}

{\makeatletter
\long\def\@makefntext#1{\parindent 1em\noindent 
\@hangfrom{\hbox to 1.8em{\hss$^{\@thefnmark}$}}#1}
\makeatother}

\def\l{\left}
\def\r{\right}
\def\p{^\prime}
\def\d{\mathrm{d}}
\newcommand{\td}[3]{\displaystyle \frac{\d^{#2}{#3}}{\d{{#1}^{#2}}}}

\newcommand{\lcmt}[2]{\l[#1,\,#2\r]}
\newcommand{\lacmt}[2]{\l\{#1,#2\r\}}
\newcommand{\cmt}[2]{[#1,\,#2]}
\newcommand{\acmt}[2]{\{#1,#2\}}

\newcommand{\itr}[1]{\mathcal I_{#1}}
\newcommand{\KBc}[0]{\textit{KBc}~}
\newcommand{\BRST}[0]{Q_\mathrm{B}}

\newcommand{\Lie}[1]{\pounds_{#1}}
\newcommand{\Lcmt}[2]{\cmt{#1}{#2}_L}

\newcommand{\Wilson}[3]{\mathcal W_{#1}(#2,#3)}
\newcommand{\wln}[2]{\mathcal W(#1,#2)}
\newcommand{\bisho}[0]{\zeta}

\newcommand{\mfd}[0]{\mathcal K}
\newcommand{\lBRST}[0]{\overleftarrow{Q}_\mathrm{B}}
\newcommand{\ITR}[2]{\itr{#1}^{(#2)}}
\newcommand{\LIE}[2]{\Lie{#1}^{(#2)}}
\def\const{constant~}

\newcommand{\identity}[1]{\mathbb I_{#1}}
\newcommand{\iitr}[0]{\mathcal I}
\newcommand{\ghost}[0]{N_{\mathrm{gh}}}
\newcommand{\mphi}[1]{\phi_{#1}}
\newcommand{\dm}[1]{\phi^\ast_{#1}}
\newcommand{\calO}{\mathcal{O}}
\newcommand{\calS}{\mathcal{S}}
\newcommand{\QB}{Q_\textrm{B}}
\newcommand{\nn}{\nonumber}
\newcommand{\tr}{\mathop{\rm Tr}}

\newcommand{\DPsi}{\Delta\Psi}
\newcommand{\Psiz}{\Psi_0}

\newcommand{\rQB}{\overleftarrow{Q}_{\rm \!B}}

\newcommand{\modif}[1]{#1}
\newenvironment{modified}[0]{}{}

\title{\hfill\parbox{3cm}{\normalsize KUNS-2861}\\[12pt]
Interior Product, Lie Derivative and Wilson Line \\
in the \KBc Subsector of Open String Field Theory}
\author{
Hiroyuki Hata\footnote{hata.hiroyuki.3@gmail.com}\quad
and\quad
Daichi Takeda\footnote{takedai@gauge.scphys.kyoto-u.ac.jp}\\[12pt]
 \textit{ Department of Physics, Kyoto University, Kyoto 606-8502, Japan}
}
\date{}
\begin{document}
\maketitle
\begin{abstract}
		The open string field theory of Witten (SFT) has a close formal similarity with Chern-Simons theory in three dimensions.
This similarity is due to the fact that the former theory has concepts corresponding to forms, exterior derivative, wedge product and integration over the manifold.
In this paper, we introduce the interior product and the Lie derivative in the \KBc subsector of SFT.
The interior product in SFT is specified by a two-component ``tangent vector" and lowers the ghost number by one (like the ordinary interior product maps a $p$-form to $(p-1)$-form).
The Lie derivative in SFT is defined as the anti-commutator of the interior product and the BRST operator.
The important property of these two operations is that they respect the \KBc algebra.

Deforming the original $(K,B,c)$ by using the Lie derivative, we can consider an infinite copies of the \KBc algebra, which we call the \KBc manifold.
As an application, we construct the Wilson line on the manifold, which could play a role in reproducing degenerate fluctuation modes around a multi-brane solution.

\end{abstract}
\newpage
\tableofcontents

\newpage

\section{Introduction}
The action of Witten's open string field theory (SFT) \cite{Witten:1985cc} has close apparent resemblances with Chern-Simons (CS) action in three dimensions.
The action of SFT is given by
\begin{align}
	S_{\mathrm{SFT}} &= \frac{1}{g^2}\int \l(\frac{1}{2}\Psi\BRST\Psi + \frac{1}{3}\Psi^3\r),
	\label{eq:witten}
\end{align}
where $\Psi$ is the string field carrying ghost number $\ghost[\Psi] = 1$ and satisfying the reality condition $\Psi^\ddag = \Psi$.
This is invariant under the infinitesimal gauge transformation
\begin{align}
	\delta\Psi = \BRST\Lambda + \cmt{\Psi}{\Lambda},\label{eq:gauge_tr}
\end{align}
and the EOM from the action is given by
\begin{align}
	\BRST\Psi + \Psi^2 = 0.\label{eq:EOM}
\end{align}
Under the finite gauge transformation
\begin{align}
	\Psi \to \Psi^V= V(\BRST + \Psi)V^{-1},
	\label{eq:Psi_gauge_tr}
\end{align}
there emerges an extra topological term:
\begin{align}
	S_\mathrm{SFT} \rightarrow S_\mathrm{SFT} - \frac{1}{6g^2}\int\bigl(V\BRST V^{-1}\bigr)^3.
\end{align}

On the other hand, the action of CS theory is given by
\begin{align}
	S_{\mathrm{CS}} = \frac{k}{2\pi}\int_{M_3}\mathrm{Tr}\l(\frac{1}{2}A\d A + \frac{1}{3}A^3\r),\label{eq:CS_action}
\end{align}
where $M_3$ is a three dimensional compact manifold, and $A = A_\mu\d x^\mu$ is anti-hermitian ($A^\dagger = -A$).
The infinitesimal gauge transformation in CS theory
\begin{align}
	\delta A = \d \lambda + \cmt{A}{\lambda},
\end{align}
keeps the action invariant, while under the finite gauge transformation
\begin{align}
	A\rightarrow A^g = g(\d + A)g^{-1},\label{eq:A_gauge_tr}
\end{align}
$S_\mathrm{CS}$ transforms as
\begin{align}
	S_\mathrm{CS} \to S_\mathrm{CS} - \frac{k}{12\pi}\int_{M_3}\mathrm{Tr}\bigl(g\d g^{-1} \bigr)^3.
\end{align}

As seen from the above, there are the following correspondences:
\begin{align}
	\ast\leftrightarrow \wedge,\qquad
	 \BRST \leftrightarrow \d ,\qquad
	\Psi\leftrightarrow A ,\qquad
	\int\leftrightarrow\int_{M_3}\!\mathrm{Tr} ,\qquad
	V\leftrightarrow g,
	\label{eq:correspondence}
\end{align}
where the star product $\ast$ and the exterior product $\wedge$ are omitted in \eqref{eq:witten} and \eqref{eq:CS_action}, respectively.
More generally, a quantity with $\ghost = p$ in SFT corresponds to a $p$-form field in CS theory.

The construction of CS theory is based on the theory of differential forms.
Besides wedge product, exterior derivative and forms, the theory of differential forms contains other two important operations; interior product and Lie derivative.
However, these concepts are not known in SFT.
The purpose of this paper is to introduce the interior product and the Lie derivative in SFT, and further to apply them to construct the Wilson line, by restricting the argument to the \KBc subsector.

In the \KBc subsector of SFT \cite{Okawa:2006vm} , all quantities are represented by $K$, $B$ and $c$, which satisfy the following (anti-)commutation relations and BRST transformation rules:
\begin{align}
	&\cmt{K}{B} = 0,\qquad
	\acmt{B}{c} = \mathbb I,\qquad
	B^2 = 0,\qquad
	c^2 = 0,\qquad\label{eq:commutators} \\
	&\BRST K = 0,\qquad
	\BRST B= K,\qquad
	\BRST c = cKc.
	\label{eq:relations}
\end{align}
Eqs.\,\eqref{eq:commutators} and \eqref{eq:relations} are called \KBc algebra (see \cite{Okawa_review} for a review).
Exact classical solutions of SFT representing the tachyon vacuum \cite{Schnabl:2005gv,Erler:2009uj} and multiple branes \cite{Murata:2011ex, Hata:2011ke,Murata:2011ep,Hata_symbol} have been constructed in the \KBc subsector.

First, we construct the interior product $\itr{X}$ in SFT, which is specified by a ``\KBc tangent vector" $X$.
This operation lowers the ghost number by $1$, corresponding to that the ordinary interior product $i_X$ maps $p$-form to $(p-1)$-form.
Besides this property, we demand on $\itr{X}$ the anti-Leibniz rule, nilpotency and the consistency with the \KBc (anti-)commutation relations \eqref{eq:commutators}.
The first two properties are natural SFT version of the properties satisfied by $i_X$.
From these requirements, we find that the operation of $\itr{X}$ on $K$, $B$ and $c$ is uniquely determined by a two-component tangent vector $X = (X^1(K),X^2(K))$ consisting of two real functions of $K$.
Then we define the Lie derivative $\Lie{X}$ by $\Lie{X} = -i\acmt{\BRST}{\itr{X}}$ in analogy with the relation $\mathcal L_X = \acmt{\d}{i_X}$ for the ordinary Lie derivative $\mathcal L_X$.
Our \KBc interior products and Lie derivatives satisfy the same kind of commutation relations as the ordinary ones, by using a suitably defined Lie bracket.

Next, by using the \KBc Lie derivative, we define the \KBc manifold.
This is the space of triads $(K(\xi),B(\xi),c(\xi))$ satisfying the same \KBc algebra \eqref{eq:commutators} and \eqref{eq:relations} and specified by a two-component real function of $K$, $\xi = (\xi^1(K),\xi^2(K))$.
A triad $(K(\xi),B(\xi),c(\xi))$ is first constructed by successively applying $(1 +\Delta s \Lie{\dot \xi(s)})$ on the original triad along a curve $\xi(s)$ with parameter $s$.
By solving the differential equation expressing this process, we find that the triad does not depend on the curve, but only on its end point.
Thus, we can consistently define the \KBc manifold.
The interior product and the Lie derivative are extended to the operations on each point on the \KBc manifold.
The string field $\Psi$ is also generalized to the ``field" $\Psi(\xi)$ on the \KBc manifold.

\begin{figure}
	 \centering
	\includegraphics[width = 4cm]{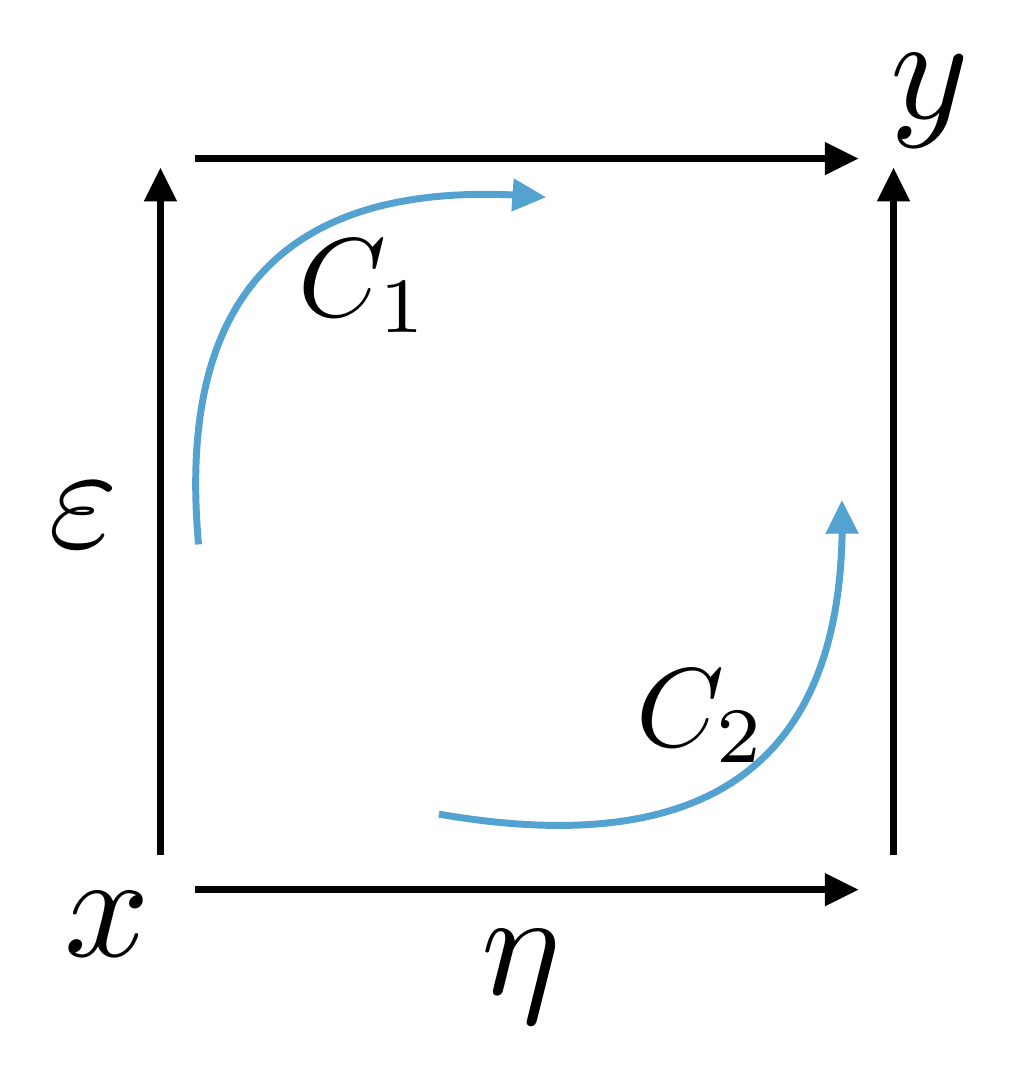}
 	\caption{The two paths $C_1$ and $C_2$ on $M_3$}
 	\label{fig:paths}
 \end{figure}

Once we established the notion of the \KBc manifold, we can introduce the Wilson line in SFT.
For explaining this, let us summarize the Wilson line in CS theory (or more generally in gauge theories).
Let $C$ be a curve $x(s)$ on $M_3$ parametrized by $s\in [a,b]$ in CS theory.
The Wilson line along $C$ is defined as 
 \begin{align}
 	W_C(x(b),x(a)) : = \mathrm{P}\exp\l(\int_C A\r) = \mathrm{P}\exp\l(\int_a^b\d s\,i_{\dot x(s)}A(x(s))\r),\label{eq:CS_Wilson}
 \end{align}
where $\mathrm{P}$ denotes the path ordering (a quantity with smaller $s$ is put more right).
If one performs a gauge transformation \eqref{eq:A_gauge_tr} on $A$, the Wilson line is transformed as
 \begin{align}
 	W_C(x(b),x(a)) \to g(x(b))\, W_C(x(b),x(a))\, g(x(a))^{-1}.\label{eq:gauge_tr_of_CS_Wilson}
 \end{align}
 Let us consider two infinitesimal paths $C_1$ and $C_2$ connecting $x$ and $y = x + \varepsilon + \eta$ with $\varepsilon$ and $\eta$ being infinitesimal constant vectors (see Fig.\,\ref{fig:paths}):
 \begin{align}
	C_1&:\,x \to x + \varepsilon  \to x + \varepsilon + \eta = y\nonumber\\
	C_2&:\,x \to x + \eta \to x + \eta + \varepsilon = y\,.
\end{align}
Then the difference between the two Wilson lines is given by the field strength $F:= \d A + A^2$ as
\begin{align}
	W_{C_1}(y,x) - W_{C_2}(y,x) = i_\eta i_\varepsilon F(x).
	\label{eq:w-w}
\end{align}
 
 In SFT, we define the Wilson line along a curve $\xi(s)$ on the \KBc manifold, by replacing $i_{\dot x(s)}A(x(s))$ in \eqref{eq:CS_Wilson} with $\itr{\dot\xi(s)}\Psi(\xi(s))$.
 This Wilson line satisfies properties similar to those in CS theory, except a number of modifications.
 The gauge transformation rule of the SFT Wilson line is different from \eqref{eq:gauge_tr_of_CS_Wilson}; there appears an extra term due to the fact that a quantity with $\ghost = 0$ is not annihilated by the operation of interior product, $\itr{X}V\neq 0$ (in contrast, we have $i_X g = 0$ in CS theory).
 By the same reason, an extra term emerges on the SFT version of the r.h.s.\ of \eqref{eq:w-w}.
 However, this extra term is missing in the special case where the Lie bracket of the two infinitesimal tangent vectors $\varepsilon$ and $\eta$ vanishes (this is the case for CS theory).
 Finally, we present the formula for the SFT Wilson line operated by $\BRST +\Psi$ which corresponds to the covariant derivative $\d + A$ in CS theory.
 We find that the Wilson line is ``almost" annihilated by the operation when $\Psi$ is on-shell.
 This formula is expected to be useful for the analysis of fluctuation modes around a multi-brane solution.
 
The organization of the rest of this paper is as follows.
We first define \KBc interior product, Lie derivative and tangent vector in section \ref{sec:itr_Lie_vec}.
In section \ref{sec:KBc}, using the \KBc Lie derivative, we introduce triads $(K(\xi),B(\xi),c(\xi))$ satisfying the \KBc algebra, which leads to the notion of the \KBc manifold.
In section \ref{sec:Wilson}, we define the Wilson line in the \KBc subsector of SFT and examine its properties.
The final section (sec.\ \ref{sec:sum}) is devoted to summary and discussions.
In appendix \ref{app:itr}, we derive the form of the \KBc interior product satisfying the conditions.
In appendix \ref{app:relation}, we present formulas for interior products and Lie derivatives used in the text.
In appendix \ref{app:multi}, we sketch a possible scenario of the emergence of degenerate excitation modes around a multi-brane solution in SFT.

\section{\KBc interior product, Lie derivative and tangent vector}\label{sec:itr_Lie_vec}

In this section, we would like to introduce the \KBc interior product and Lie derivative.
Usually, we need the notions of manifold and tangent vector on it before introducing these operations.
However, since such notions are not known in SFT, we adopt a heuristic approach here.
Namely, we first construct the \KBc interior product by imposing suitable conditions on it.
This naturally induces the \KBc tangent vector, and further the \KBc manifold.
%
In the following, all quantities are in the sliver frame \cite{Rastelli:2000iu},
 and we assume that they consist only of $K$, $B$ and $c$.

We introduce the \KBc interior product $\itr{X}$ as a linear operation on the \KBc subsector which lowers the ghost number by one and satisfies the following four conditions:
\begin{enumerate}
	\item Anti-Leibniz rule
		\begin{align}
			\itr{X}(\mathcal A\mathcal B) = (\itr{X}\mathcal A)\mathcal B + (-1)^{|\mathcal A|}\mathcal A(\itr{X}\mathcal B),\label{eq:Leibniz}
		\end{align}
	\item Double-conjugation property
		\begin{align}
			(\itr{X}\mathcal A)^\ddag = -(-1)^{|\mathcal A|}\itr{X}\mathcal A^\ddag,\label{eq:doule_conjugation}
		\end{align}
	\item Nilpotency
		\begin{align}
			(\itr{X})^2 = 0,\label{eq:nilpotency}
		\end{align}
	\item Consistency with \KBc (anti-)commutation relations \eqref{eq:commutators}.
\end{enumerate}
Here $\mathcal A$ and $\mathcal B$ are any quantities consisting of $K$, $B$ and $c$, and
\begin{align}
	(-1)^{|\mathcal A|} := 
	\begin{cases}
		1 & \mbox{($\mathcal A$ is Grasmann-even})\\
		-1 & \mbox{($\mathcal A$ is Grasmann-odd})
	\end{cases}.
\end{align}
At this stage, we have assumed that each \KBc interior product $\itr{X}$ is specified by some quantity $X$.
The properties 1 and 3 are the SFT version of the properties satisfied by the ordinary interior product.
Note that the first property implies, in particular, that $\itr{X}\mathbb I = 0$.
The property 2 is the hermiticity of $\itr{X}$, which is also satisfied by $\BRST$.
The fourth property means that the operation of $\itr{X}$ on both hand sides of each relation in \eqref{eq:commutators} keeps the equality. 
For example,
\begin{align}
	\itr{X}(\{B,c\}) = \itr{X}\mathbb I = 0,\qquad
	\itr{X}c^2 = \itr{X}0 = 0,
\end{align}
should hold for $\{B,c\} = \mathbb I$ and $c^2 = 0$, respectively.

As shown in appendix \ref{app:itr}, the most general form of the operation of $\itr{X}$ on $K$, $B$ and $c$ is given by
\begin{align}
	\itr{X}K = iBX^1,\qquad
	\itr{X}B = 0,\qquad
	\itr{X}c = \frac{X^2}{K} + \lcmt{\frac{X^2}{K}}{Bc},\label{eq:def_interior}
\end{align}
where $X = \l(X^1(K),X^2(K)\r)$ is a two-component real function of $K$, and is called \KBc tangent vector.\footnote{
\modif{Readers may wonder why $X^2$ in \eqref{eq:def_interior} is divided by $K$.
The reason is that, in order for the solution of \eqref{eq:K_sB_sc_s} to be path-independent, we should regard $X^2$ and not $X^2/K$ as the second component of the \KBc tangent vector.}
}
\modif{\KBc tangent vector $X$ is supposed to correspond to tangent vector at a point on $M_3$ in CS theory.
This will be generalized to \KBc vector field later.}

Different \KBc tangent vectors give different \KBc interior products.
For any two tangent vectors\footnote{Hereafter, we often omit ``\textit{KBc}" which distinguishes the \KBc version from that in the theory of differential forms.}
 $X$ and $Y$, $\itr{X}$ and $\itr{Y}$ anti-commute with each other:
\begin{align}
	\acmt{\itr{X}}{\itr{Y}} = 0.\label{eq:itr_anti_commute}
\end{align}
This is shown as follows.
First, $\acmt{\itr{X}}{\itr{Y}}\mathcal A = 0$ holds for $\mathcal A = K,\,B,\,c$ since we have
\begin{align}
 	\itr{X}\itr{Y}\mathcal A = 0\quad(\mathrm{for}\ \mathcal A = K,\,B,\,c).
\end{align}
Next, $\acmt{\itr{X}}{\itr{Y}}\mathcal A = 0$ for a generic $\mathcal A$ which is a sum of products of $K$, $B$ and $c$ is shown by induction by using the identity following from the anti-Leibniz rule \eqref{eq:Leibniz}:
\begin{align}
	\acmt{\itr{X}}{\itr{Y}}(\mathcal{AB})
	&= (\itr{X}\itr{Y}\mathcal A)\mathcal B + \mathcal A\itr{X}\itr{Y}\mathcal B\nonumber\\
	&\qquad +(-1)^{|\mathcal A|}\l[(\itr{X} \mathcal A)\itr{Y}\mathcal B - (\itr{Y}\mathcal A)\itr{X}\mathcal B\r] + (X\rightleftarrows Y)\nonumber\\
	&= (\acmt{\itr{X}}{\itr{Y}}\mathcal A)\mathcal B + \mathcal A(\acmt{\itr{X}}{\itr{Y}}\mathcal B).
\end{align} 

Another good property is 
\begin{align}
	\itr{\alpha X + \beta Y} = \alpha\itr{X} + \beta\itr{Y},\label{eq:itr_index_linear}
\end{align}
where $\alpha$ and $\beta$ are real numbers and
\begin{align}
 	\alpha X + \beta Y = (\alpha X^1 + \beta Y^1,\alpha X^2 + \beta Y^2).
\end{align}
This follows from \eqref{eq:def_interior} and the anti-Leibniz rule \eqref{eq:Leibniz}.

There is a critical difference between the \KBc interior product $\itr{X}$ and the ordinary one $i_X$; the latter annihilates the $0$-forms, while $\itr{X}$ does not annihilate quantities with $\ghost = 0$.
Instead, we have $\itr{X}\mathcal O = 0$ for any $\mathcal O$ with $\ghost[\mathcal O] = -1$ since there is no quantity consisting only of $K$, $B$ and $c$ and carrying $\ghost \leq -2$.

Now we define the \KBc Lie derivative $\Lie{X}$ as
\begin{align}
	\Lie{X} := -i\acmt{\BRST}{\itr{X}},
\end{align}
which carries no ghost number.
This is of the same form as the ordinary Lie derivative $\mathcal L_X = \acmt{\d}{i_X}$, except for the phase factor $-i$. 
From the double-conjugation property \eqref{eq:doule_conjugation} of $\itr{X}$ and that of $\BRST$, we see that $\Lie{X}$ enjoys
\begin{align}
	(\Lie{X}\mathcal A)^\ddag = \Lie{X}\mathcal A^\ddag.
\end{align}
Our \KBc Lie derivative $\Lie{X}$ shares the following properties with the ordinary one $\mathcal L_X$ (with of course the replacement $\d \rightarrow \BRST)$:
\begin{align}
 	&\cmt{\Lie{X}}{\BRST} = 0,\qquad\cmt{\Lie{X}}{\itr{Y}} = \cmt{\itr{X}}{\Lie{Y}} = - \cmt{\Lie{Y}}{\itr{X}},\nonumber\\
 	&\Lie{X}(\mathcal{AB}) = (\Lie{X}\mathcal A)\mathcal B + \mathcal A(\Lie{X}\mathcal B),\qquad
 	\Lie{\alpha X + \beta Y} = \alpha\Lie{X} + \beta\Lie{Y}.
\end{align}

The concrete action of $\Lie{X}$ on $K$, $B$ and $c$ are as follows:
\begin{align}
	\Lie{X}K &= KX^1,\qquad
	\Lie{X}B = BX^1,\qquad\Lie{X}c = -cX^1Bc - i\cmt{X^2}{c}.\label{eq:Lie_action}
\end{align}
An important property of $\Lie{X}$ is that the transformation with an infinitesimal constant $\varepsilon$,
\begin{align}
 	(K,B,c)\to (K\p,B\p,c\p) = (1+\varepsilon\Lie{X})(K,B,c),\label{eq:KBctoKBc}
\end{align}
 keeps the \KBc algebra, namely $(K\p,B\p,c\p)$ satisfies \eqref{eq:commutators} and \eqref{eq:relations} to $O(\varepsilon)$.
 This is because $\itr{X}$ and $\BRST$ keep \eqref{eq:commutators} and $\Lie{X}$ commutes with $\BRST$.
For this reason, if $\Psi$ is a solution to the EOM \eqref{eq:EOM}, $(1 +\varepsilon \Lie{X})\Psi$ is also a solution to $O(\varepsilon)$, because only the \KBc algebra is used when one shows that $\Psi$ consisting of $K$, $B$ and $c$ is a solution.
In fact, $\varepsilon \Lie{X}\Psi$ for a solution $\Psi$ is equal to the infinitesimal gauge transformation \eqref{eq:gauge_tr} with $\Lambda = -i\varepsilon \itr{X}\Psi$:
\begin{align}
 	\Lie{X}\Psi &= - i\BRST\itr{X}\Psi - i\itr{X}\BRST\Psi = - i\BRST\itr{X}\Psi + i\itr{X}\Psi^2\nonumber\\
 	&= - i\BRST\itr{X}\Psi + \cmt{\Psi}{-i\itr{X}\Psi}.
\end{align}

The commutator of the ordinary Lie derivative $\mathcal L_{X}$ and the interior product $i_Y$ is again an interior product: $\cmt{\mathcal L_X}{i_Y} = i_{\cmt{X}{Y}}$.
This is also the case for the present $\Lie{X}$ and $\itr{Y}$ (with the replacement of the commutator $\cmt{X}{Y}$ by a suitable one).
In order to see this, let us first consider the action of $\cmt{\Lie{X}}{\itr{Y}}$ on $K$, $B$ and $c$:
\begin{align}
	\cmt{\Lie{X}}{\itr{Y}}K &= iB\l(X^1K\partial Y^1 - Y^1K\partial X^1\r),\\
	\cmt{\Lie{X}}{\itr{Y}}B &= 0,\\
	\cmt{\Lie{X}}{\itr{Y}}c &= X^1K\partial Y^2 - Y^1K\partial X^2+ \lcmt{X^1K\partial Y^2 - Y^1K\partial X^2}{Bc}.
\end{align}
with $\partial: = \partial/\partial K$.
Therefore defining $\Lcmt{X}{Y}$ as 
\begin{align}
	\Lcmt{X}{Y} := \l(X^1K\partial Y^1 - Y^1K\partial X^1,X^1K\partial Y^2 - Y^1K\partial X^2\r),\label{eq:Lcmt}
\end{align}
we find that the relation
\begin{align}
	\cmt{\Lie{X}}{\itr{Y}} = \itr{\Lcmt{X}{Y}}\label{eq:Lie_itr_cmt},
\end{align}
holds at least when the both hand sides act on $K$, $B$ and $c$.
Then using that $\cmt{\Lie{X}}{\itr{Y}}$ satisfies the same anti-Leibniz rule as \eqref{eq:Leibniz} for $\itr{X}$, we see that \eqref{eq:Lie_itr_cmt} holds against
any quantity consisting of $K$, $B$, and $c$.
In addition, because $\Lie{X}$ commutes with $\BRST$, one also gets
\begin{align}
	\cmt{\Lie{X}}{\Lie{Y}} &= -i\cmt{\Lie{X}}{\BRST\itr{Y} + \itr{Y}\BRST}
	= -i\BRST\cmt{\Lie{X}}{\itr{Y}} - i\cmt{\Lie{X}}{\itr{Y}}\BRST\nonumber\\
	&= -i(\BRST\itr{\Lcmt{X}{Y}} - i\itr{\Lcmt{X}{Y}}\BRST)= \Lie{\Lcmt{X}{Y}},\label{eq:Lie_cmt}
\end{align}
as expected.

The bracket $\Lcmt{X}{Y}$ is a Lie bracket, namely, it satisfies the bilinearity, the anti-symmetry
\begin{align}
	\Lcmt{X}{Y} = - \Lcmt{Y}{X},
\end{align}
and the Jacobi identity
\begin{align}
	\Lcmt{X}{\Lcmt{Y}{Z}} + \Lcmt{Y}{\Lcmt{Z}{X}} + \Lcmt{Z}{\Lcmt{X}{Y}} = 0.\label{eq:Jacobi}
\end{align}
In fact, by introducing\footnote{
$\hat X$ given by \eqref{eq:hat} can be replaced by $\hat X = X^1K\partial + \lambda X^2$ with any complex number $\lambda$.
}
\begin{align}
	\hat X = X^1K\partial + X^2,\label{eq:hat}
\end{align}
it is not hard to show that
\begin{align}
	\cmt{\hat X}{\hat Y} = (\Lcmt{X}{Y})^1K\partial + (\Lcmt{X}{Y})^2 = \widehat{\Lcmt{X}{Y}},
\end{align}
so $\Lcmt{X}{Y}$ is reduced to a simple commutator $\cmt{\hat X}{\hat Y}$.
This implies that $\Lcmt{X}{Y}$ is a Lie bracket.

We can simplify the expression of the Lie bracket \eqref{eq:Lcmt} as
\begin{align}
	\Lcmt{X}{Y} = X^1K\partial Y - Y^1K\partial X = \Lie{X}Y - \Lie{Y}X. \label{eq:simple_Lcmt}
\end{align}
This expression also holds for the hat version\footnote{
Defining $V_X := X^1K\partial$, \eqref{eq:2.29} is rewritten as
$\cmt{V_X + X^2}{V_Y + Y^2} = \cmt{V_X}{V_Y}  + \Lie{X}Y^2 - \Lie{Y}X^2$.
This is apparently of the form of Courant bracket for the direct sum of tangent vector and $0$-form.
This resemblance may give a clue for deeper understanding of the \KBc manifold introduced in the next section.
}
\begin{align}
	\cmt{\hat X}{\hat Y} = \Lie{X}\hat Y - \Lie{Y}\hat X,\label{eq:2.29}
\end{align}
because $\Lie{X} = X^1K\partial$ for any function of $K$.

\section{\KBc manifold}\label{sec:KBc}
\subsection{The construction}
In order to introduce the Wilson line in SFT, first of all, we should establish the notion of a manifold, which corresponds to $M_3$ in CS theory.
We shall show that each point on the manifold corresponds to a certain triad of $(K(\xi),B(\xi),c(\xi))$ specified by $\xi = (\xi^1(K),\xi^2(K))$, which is a two-component real function of  $K$.
Each triad $(K(\xi),B(\xi),c(\xi))$ satisfies the same \KBc algebra as the original one $(K,B,c)$.

Let $\xi_s = (\xi_s^1(K),\xi_s^2(K))$ be a two-component real function of $K$ parametrized by a real variable $s$, and we set $\xi_{s=0} = 0$.
For this $\xi_s$, let us consider the triad $(K_s,B_s,c_s)$ determined by the following differential equation and initial condition:
\begin{align}
	\td{s}{}{}(K_s,B_s,c_s) &= \Lie{\dot\xi_s}(K_s,B_s,c_s),\qquad
	(K_0,B_0,c_0) = (K,B,c),
	\label{eq:K_sB_sc_s}
\end{align}
where the Lie derivative $\Lie{\dot\xi_s}$ is defined to be given by \eqref{eq:Lie_action} with $(K,B,c)$ replaced by $(K_s,B_s,c_s)$, and the dot denotes the $s$ derivative.
Concretely, \eqref{eq:K_sB_sc_s} reads
\begin{align}
	\dot K_s &= \dot\xi_s^1 K_s,\label{eq:K_diff}\\
	\dot B_s &= \dot\xi_s^1 B_s,\label{eq:B_diff}\\
	\dot c_s &= -c_s\dot\xi_s^1 B_sc_s - i \cmt{\dot\xi^2_s}{c_s}.\label{eq:c_diff}
\end{align}
Since the Lie derivative preserves the \KBc algebra, if $(K_s,B_s,c_s)$ satisfies \KBc algebra, then
$$(K_s + \delta s\dot K_s,B_s + \delta s\dot B_s,c_s +  \delta s\,\dot c_s)$$
also satisfies it to $O(\delta s)$.
Therefore, $(K_s,B_s,c_s)$ determined by \eqref{eq:K_sB_sc_s} satisfies the \KBc algebra for any $s$.

Let us solve the differential equations \eqref{eq:K_diff} -- \eqref{eq:c_diff}.
Because $K$ commutes with $B$, \eqref{eq:K_diff} and \eqref{eq:B_diff} can be easily solved to give
\begin{align}
	K_s = e^{\xi^1_s}K,\qquad B_s = e^{\xi^1_s}B.\label{eq:KB_s}
\end{align}
In order to solve \eqref{eq:c_diff}, we first consider the differential equation for $\sigma_s := B_sc_s$, which is obtained from \eqref{eq:B_diff} and \eqref{eq:c_diff}:
\begin{align}
	\dot\sigma_s = -i\cmt{\dot\xi^2_s}{\sigma_s}.
\end{align}
This gives
\begin{align}
	\sigma_s = e^{-i\xi^2_s}\sigma_0\, e^{i\xi^2_s} = e^{-i\xi^2_s}Bc\,e^{i\xi^2_s}.\label{eq:sigma_s}
\end{align}
Substituting \eqref{eq:sigma_s} back to \eqref{eq:c_diff}, we obtain
\begin{align}
 	\dot c_s = -c_s\dot\xi_s^1e^{-i\xi^2_s}Bc \,e^{i\xi^2_s} - i\cmt{\dot\xi^2_s}{c_s},
\end{align}
which reduces to a simpler equation for $\tilde c_s := e^{i\xi^2_s}c_s e^{-i\xi^2_s}$:
\begin{align}
	\dot{\tilde{c}}_s = -\tilde c_s \dot\xi_s^1Bc.
\end{align}
This can be solved to give
\begin{align}
 	\tilde c_s  = c\,e^{-\xi^1_s}Bc,
\end{align}
and we finally obtain
\begin{align}
	c_s = e^{-i\xi^2_s}c\,e^{-\xi_s^1}Bc\,e^{i\xi^2_s}.\label{eq:c_s}
\end{align}

From \eqref{eq:KB_s} and \eqref{eq:c_s}, we see that the solution $(K_s,B_s,c_s)$ depends only on $\xi_s$ and not on the intermediate $\xi_t$ and $\dot \xi_t$ for $0\le t <s$:
the solution is completely specified by the end point.
This fact leads us to define the \KBc manifold $\mfd$ as follows:
\begin{itemize}
	\item $\mfd$ contains all the solutions of \eqref{eq:K_sB_sc_s} as the points on $\mfd$.
	\item Each point has the expression
		\begin{align}
			K(\xi) &= e^{\xi^1}K,\quad
			B(\xi) = e^{\xi^1}B,\quad
			c(\xi) = e^{-i\xi^2}c\,e^{-\xi^1}Bc\,e^{i\xi^2},
			\label{eq:KBc_xi}
		\end{align}
		in terms of a two-component real function $\xi = (\xi^1(K),\xi^2(K))$.
		We regard $\xi$ as the coordinate of $\mfd$ (or often as the point on $\mfd$).\footnote{
		In the special case of $\xi^2 = 0$,  \eqref{eq:KBc_xi} is the EMNT transformation \cite{Erler:2012dz,Masuda:2012kt,erler,Hata:2012cy}.
		In the EMNT transformation, $e^{\xi^1(K)}$ is expressed as an arbitrary real function $g(K)$. However, our restriction of $g(K)$ to $e^{\xi^1(K)}$ seems more natural since it keeps the property $K(\xi)\geq 0$.
		}
\end{itemize}
In short, the \KBc manifold $\mfd$ is regarded as the space of two-component real functions of $K$.
In order for the triad $(K(\xi),B(\xi),c(\xi))$ also to span the \KBc subsector, it is necessary and sufficient that the original triad $(K,B,c)$ is expressed by this triad.
By tracing the curve $C$ in the reverse direction from its end point $\xi$ to the origin, of course, the original triad is expressed as
\begin{align}
	K = e^{-\xi^1}K(\xi),\qquad
	B = e^{-\xi^1}B(\xi),\qquad
	c = e^{i\xi^2}c(\xi)e^{\xi^1}B(\xi)c(\xi)\, e^{-i\xi^2}.\label{eq:inverse}
\end{align}
However, \eqref{eq:inverse} does not give the original triad $(K,B,c)$ in terms of $(K(\xi),B(\xi),c(\xi))$, since $e^{\pm \xi^1}$ and $e^{\pm i\xi^2}$ on the r.h.s.\ are functions of the original $K$.
Therefore, by restricting ourselves to the coordinate patch such that the map $K\mapsto K(\xi) = e^{\xi^1(K)}K$ is one-to-one, any quantity in the \modif{\KBc subsector} can be represented by any triad on $\mfd$.

The argument in the previous section about the tangent vector, interior product and Lie derivative for the original triad can be generalized to $\mfd$.
\modif{Now we can define a \KBc vector field  $X(\xi)$ on $\mfd$, which returns a tangent vector for each point $\xi\in\mfd$, like a vector field in CS theory.}
An example is
\begin{align}
	X(\xi) = \l(\xi^1 + (\xi^2)^2 + 3K,\,e^{\xi^1} + K^2 + 1\r).\label{eq:example}
\end{align}
As given in \eqref{eq:example}, $X(\xi)$ may have dependences on $K=K(0)$ not through $\xi^1(K)$ and $\xi^2(K)$.
The special class of \modif{vector fields} without the dependence on $\xi$, for example,
\begin{align}
	X(\xi) = (K^2+2,\, e^K),\label{eq:const_example}
\end{align}
\modif{are called \const vector fields, or \const vectors for short.}

The interior product defined in \eqref{eq:def_interior} for the original triad can be extended to any point $(K(\xi),B(\xi),c(\xi))$ in $\mfd$.
Denoting the interior product at $\xi$ by $\ITR{X}{\xi}$, which coincides with $\itr{X}$ at $\xi = 0$, its operation is given by
\begin{align}
	\ITR{X}{\xi}K(\xi) &= iB(\xi)X^1(\xi),\qquad
	\ITR{X}{\xi}B(\xi) = 0,\nonumber\\
	\ITR{X}{\xi}c(\xi) &= \frac{X^2(\xi)}{K(\xi)} + \lcmt{\frac{X^2(\xi)}{K(\xi)}}{B(\xi)c(\xi)}.\label{eq:xi_itr}
\end{align}
The corresponding Lie derivative $\LIE{X}{\xi}$ is defined as
\begin{align}
	\LIE{X}{\xi} = -i\bigl\{\BRST,\ITR{X}{\xi}\bigr\},
\end{align}
and its operation on $K(\xi)$, $B(\xi)$ and $c(\xi)$ is given by
\footnote{
The relation between the interior products (and the Lie derivatives) at $\xi$ and at the origin is given in appendix \ref{app:relation}.
}
\begin{align}
	\LIE{X}{\xi}K(\xi) &= K(\xi)X^1(\xi),\qquad
	\LIE{X}\xi B(\xi) = B(\xi)X^1(\xi),\quad\nonumber\\
	\LIE{X}\xi c(\xi) &= -c(\xi)X^1(\xi)B(\xi)c(\xi) - i\cmt{X^2(\xi)}{c(\xi)}.\label{eq:xi_Lie_action}
\end{align}
The other properties \eqref{eq:Lie_itr_cmt} and \eqref{eq:Lie_cmt} also hold for the present case, with the Lie bracket
\begin{align}
	\Lcmt{X}{Y}(\xi) = \LIE{X}{\xi}Y(\xi) - \LIE{Y}{\xi}X(\xi).\label{eq:L-L}
\end{align}
In the rest of this paper, we often omit the index ${(\xi)}$ on $\ITR{X}{\xi}$ and $\LIE{X}{\xi}$ if there is no ambiguity.

\begin{modified}
	Finally, let us clarify the correspondence between $M_3$ and $\mfd$ by referring to their dimensionality.
	In CS theory, $M_3$ is a manifold of real dimension three, and the coordinate $x$ has three real components $(x^1,x^2,x^3)$.
	A tangent vector at $x \in M_3$ is also expressed as $(v^1,v^2,v^3)$ and a vector field as $(v^1(x),v^2(x),v^3(x))$.
	Our \KBc manifold is described by the coordinate $\xi = (\xi^1(K),\xi^2(K))$, which has two components that are real functions of $K$.
	Assuming that they can be Taylor-expanded as $\xi^i(K) = \sum_n a_n^{(i)} K^n$, the dimension of $\mfd$ is countably infinite.
	A tangent vector is expressed as $(X^1,X^2) = (\sum_n X^{(1)}_nK^n,\sum_n X^{(2)}_nK^n)$ and a vector field as $(X^1(\xi),X^2(\xi)) = (\sum_n X^{(1)}_n(\xi)K^n,\sum_n X^{(2)}_n(\xi)K^n)$, where each $X_n^{(i)}(\xi)$ returns a real number for each point $\xi\in\mfd$.
\end{modified}


\subsection{String field on $\mfd$}\label{sec:SFT_in_K}
In the Introduction, we discussed the correspondence between SFT and CS theory.
Now that the \KBc manifold has been introduced, it is more natural to modify the relation $A\leftrightarrow \Psi$ to
\begin{align}
	A(x)\leftrightarrow \Psi(\xi).
\end{align}
Here $\Psi(\xi)$ is obtained by the replacement
\begin{align}
 	(K,B,c) \to (K(\xi),B(\xi),c(\xi))\label{eq:replacement}
\end{align}
for all $(K,B,c)$ in $\Psi$.
In general, when an operator $\mathcal O$ is given by $(K,B,c)$, the new operator obtained by the replacement \eqref{eq:replacement} is denoted by $\mathcal O|_{0\to\xi}$.\footnote{
It is possible to consider $\Psi(\xi)$ having the $\xi$ dependences not through $(K(\xi),B(\xi),c(\xi))$, like the \modif{vector field} \eqref{eq:example}.
However, we restrict ourselves only to $\Psi(\xi) = \Psi|_{0\to \xi}$ for which \eqref{eq:Psi_xi} holds.
}

Under the translation $\xi \to \xi + \delta\xi$ with $\delta\xi$ being an infinitesimal constant vector, $\Psi(\xi)$ transforms as
\begin{align}
	\Psi(\xi) \to \Psi(\xi + \delta\xi) = (1 + \Lie{\delta\xi})\Psi(\xi).\label{eq:Psi_xi}
\end{align}
This is of the same form as in CS theory.
Let $\varepsilon$ be an infinitesimal constant vector on $M_3$.
Then by the translation $x \to x + \varepsilon$, the gauge field 1-form $A(x)$ transforms as
\begin{align}
	A(x) \to A(x+\varepsilon) = (1 + \mathcal L_\varepsilon)A(x),
\end{align}
because $\partial_\mu\varepsilon^\nu = 0$.
Note that, from \eqref{eq:Psi_xi} and the discussion just below \eqref{eq:KBctoKBc}, $\Psi(\xi)$ is automatically on-shell if $\Psi$ is on-shell.

For the finite gauge transformation at $\xi = 0$,
\begin{align}
   \Psi^V = V(\BRST + \Psi)V^{-1},
\end{align}
the following relation holds for $\Psi^V(\xi) = \Psi^V|_{0\to \xi}$ and $V(\xi) = V|_{0\to\xi}$:
\begin{align}
  \Psi^V(\xi) = V(\xi)(\BRST + \Psi(\xi))V^{-1}(\xi).
  \label{eq:Psi_V}
\end{align}
This is because $\Psi^V$ is also another string field at $\xi = 0$.

\section{Wilson lines in SFT}\label{sec:Wilson}
\subsection{The definition}
The infinitesimal version of \eqref{eq:CS_Wilson}, which we call Wilson link, is
\begin{align}
	W(x + \varepsilon,x)  = 1 + i_\varepsilon A(x)= 1 + \varepsilon^\mu A_\mu(x) .\label{eq:CSlink}
\end{align}
Taking the hermitian conjugate, we have to $O(\varepsilon)$
\begin{align}
 	W(x+\varepsilon,x)^\dagger = 1 - \varepsilon^\mu A_\mu(x)
 	 = 1 - \varepsilon^\mu A_\mu(x+\varepsilon)
 	  = W(x,x + \varepsilon).\label{eq:wCS_dagger}
\end{align}
In SFT, let us define the Wilson link specified by an infinitesimal constant vector $\bisho$ by
\begin{align}
	\wln{\xi + \bisho}{\xi} := 1 + i\itr{\bisho}\Psi(\xi).
	\label{eq:wln}
\end{align}
Note that we have $\mathcal W(\xi + \bisho,\xi)^\ddag = \wln{\xi}{\xi + \bisho}$ by the same procedure as \eqref{eq:wCS_dagger}, which is due to \eqref{eq:doule_conjugation} and the reality condition of $\Psi$; $\Psi^\ddag = \Psi$.

In CS theory, the Wilson link is gauge-transformed as \eqref{eq:gauge_tr_of_CS_Wilson}:
\begin{align}
 	W(x + \varepsilon,x) \to g(x+\varepsilon)W(x + \varepsilon,x)g(x)^{-1}.
\end{align}
Unfortunately, the same gauge transformation rule does not hold for the SFT Wilson link.
In order to see this, we consider the case of $\wln{\bisho}{0}$ for simplicity, but the result for $\wln{\xi+\bisho}{\xi}$ can be obtained by the replacement $V\to V(\xi)$, $\Psi\to\Psi(\xi)$ in the following.
Under the gauge transformation of $\Psi$ given by \eqref{eq:Psi_gauge_tr}, $\wln{\bisho}{0}$ transforms as 
\def\1{\itr{\bisho}}
\def\2{^{-1}}
\def\3{\Lie{\bisho}}
\begin{align}
	\wln{\bisho}{0} \to& 1 + i\1\l[V(\BRST + \Psi)V\2\r]= 1 - i\1\l[(\BRST V)V\2\r] + i\1\l[V\Psi V\2\r]\nonumber\\
	&=  [(1 + \3)V]V\2 + i(\BRST\1 V)V\2 + i (\BRST V)\1 V\2\nonumber\\
			&	\hspace{42pt} +  i(\1 V)\Psi V\2 + iV(\1 \Psi )V\2 - iV\Psi(\1 V\2) \nonumber\\
	&= [(1 + \3)V]V\2 + iV\l[\BRST\l(V\2\1 V\r)\r]V\2\nonumber\\
			& \hspace{42pt} + iV\lacmt{\Psi}{V\2\1 V} V\2 + iV(\1\Psi) V\2\nonumber\\
	&= \bigl((1 + \Lie{\zeta})V\bigr)\l[1 + i\itr{\zeta}\Psi + iQ_\Psi (V\2\1 V)\r]V^{-1},\label{eq:W_gauge}
\end{align}
up to $O(\bisho^2)$. In this derivation, we have used $V\BRST V^{-1} = - (\BRST V)V^{-1}$ at the first equality, and $V\itr{\bisho}V^{-1} = - (\itr{\bisho} V)V^{-1}$ at the third one.
The operator $Q_\Psi$ is the BRST operator on the background $\Psi$ defined as
\begin{align}
	Q_\Psi \mathcal A :=  \BRST\mathcal A + \Psi\mathcal A - (-1)^{|\mathcal A|}\mathcal A\Psi.
	\label{eq:BRST_around}
\end{align}
Besides the expected term $(1 + \Lie{\zeta}) V = V(\zeta)$ in the last expression of \eqref{eq:W_gauge}, there has emerged the extra term $i Q_\Psi (V\2\1 V)$ due to the fact that $\itr{\bisho} V$ does not vanish in general.
In CS theory, the gauge parameter $g(x)$ is a $0$-form, so it vanishes by the action of the interior product.

Let us proceed to the Wilson line.
Let $C$ be a curve $\xi(s)$ on $\mfd$.
By analogy with \eqref{eq:CS_Wilson}, we define the SFT Wilson line as
\begin{align}
	\Wilson{C}{\xi(b)}{\xi(a)} = \mathrm{P}\exp\l(i\int_a^b\d s\,\itr{\dot\xi(s)}\Psi(\xi(s))\r).
	\label{eq:wSFT}
\end{align}
Setting $\xi(a) = \xi$ and $\xi(b) =  \xi + \bisho$, this is reduced to \eqref{eq:wln}.
Regarding the Wilson line as a product of Wilson links and using the gauge transformation rule \eqref{eq:W_gauge}, we find that the gauge transformation rule of our Wilson line \eqref{eq:wSFT} is as follows:
\begin{align}
 	&\Wilson{C}{\xi(b)}{\xi(a)} \rightarrow \nonumber\\
 	&V(\xi(b))\,
 	\mathrm{P}\exp\l[
 		i\int_a^b\d s\Bigr(\itr{\dot\xi(s)}\Psi(\xi(s)) 
 		+ Q_{\Psi(\xi(s))}\bigl(V(\xi(s))^{-1}\itr{\dot\xi(s)}V(\xi(s))\bigr)\Bigr)
 	\r]
 	V(\xi(a))^{-1}.
 	\label{eq:finite_W_gauge}
\end{align}
As given in \eqref{eq:finite_W_gauge}, there appears an extra term in the exponent.
However, as we shall see in the next subsection, some nice properties still hold for the present Wilson line, though they are deformed from the corresponding ones in CS theory.

\subsection{Some properties}\label{subsec:properties}
First, let us consider the SFT counterpart of \eqref{eq:w-w}.
Let $C_1$ and $C_2$ be two infinitesimal paths on $\mfd$ connecting $\xi$ and $\xi\p$:
\begin{align}
	C_1&: \xi \to \xi + \zeta\to \xi + \zeta + \eta = \xi\p\nonumber\\
	C_2&: \xi \to \xi + \eta \to \xi + \eta + \zeta = \xi\p.
\end{align}
Here $\zeta$ and $\eta$ are infinitesimal \const tangent vectors.
For notational simplicity, we abbreviate $\Psi(\xi)$ and $\Psi(\xi+\zeta)$ as  $\Psi$ and $\Psi_{\zeta}$, respectively.
Let us calculate the difference between $\Wilson{C_1}{\xi\p}{\xi}$ and $\Wilson{C_2}{\xi\p}{\xi}$
\def\itrv{\itr{\zeta}}
\def\itre{\itr{\eta}}
\begin{align}
	\Wilson{C_1}{\xi\p}{\xi} - \Wilson{C_2}{\xi\p}{\xi} &= (1 + i\itre\Psi_\zeta)(1 + i\itrv\Psi)- (\eta\rightleftarrows \zeta).
\end{align}
To be precise, $\itre\Psi_\zeta$ should be written as $\ITR{\eta}{\xi+\zeta}\Psi(\xi+\zeta)$, so by using \eqref{eq:itr_to_itr}\footnote{
In appendix \ref{app:relation}, we derive the relation between interior products (and Lie derivatives) at different points on the \KBc manifold.
} and \eqref{eq:Psi_xi}, $\itre\Psi_\zeta$ is expanded as follows:
\begin{align}
 	\itre^{(\xi+\zeta)}\Psi_\zeta &= \ITR{\eta}\xi\Psi_\zeta - \itr{\LIE{\eta}{\xi}\zeta}^{(\xi)}\Psi_\zeta = \itr{\eta}^{(\xi)}\Psi + \itr{\eta}^{(\xi)}\Lie{\zeta}^{(\xi)}\Psi - \itr{\LIE{\eta}{\xi}\zeta}^{(\xi)}\Psi.
\end{align}
Omitting the superscript ${(\xi)}$, we get
\begin{align}
	&\Wilson{C_1}{\xi\p}{\xi} - \Wilson{C_2}{\xi\p}{\xi}= i(\itr{\eta}\Lie{\zeta} - \itr{\zeta}\Lie{\eta})\Psi-\cmt{\itr{\eta}\Psi}{\itr{\zeta}\Psi} - i\itr{\Lcmt{\eta}{\zeta}}\Psi,
	\label{eq:w-w_med}
\end{align}
by using \eqref{eq:L-L}.
For the first and second terms on the r.h.s., the following formulas hold:
\begin{align}
 	i(\itr{\eta}\Lie{\zeta} - \itr{\zeta}\Lie{\eta}) &= \cmt{\itr{\eta}\itr{\zeta}}{\BRST} + i\itr{\Lcmt{\eta}{\zeta}},\\
 	\cmt{\itr{\eta}\Psi}{\itr{\zeta}\Psi} &= -\itr{\eta}\itr{\zeta}\Psi^2 + \acmt{\Psi}{\itr{\eta}\itr{\zeta}\Psi}.
\end{align}
We finally obtain
\begin{align}
	\Wilson{C_1}{\xi\p}{\xi} - \Wilson{C_2}{\xi\p}{\xi} &= -\itr{\eta}\itr{\zeta}(\BRST\Psi + \Psi^2) + Q_\Psi(\itr{\zeta}\itr{\eta}\Psi).\label{eq:W-W} 
\end{align}
On the r.h.s.\ of this formula, the first term corresponds to the r.h.s.\ of \eqref{eq:w-w}, but there exists an additional term in SFT.

In the special case of $\zeta = (0,h)$ and $\eta = (0,g)$, the last term of \eqref{eq:W-W} vanishes.
This is shown as follows.
In the notation introduced in \cite{Hata_symbol}, $\Psi$ is generally expressed as $\Psi_{13} = F_{123}c_{12}(Bc)_{23}$ with $F_{123} = F(K_1,K_2,K_3)$.\footnote{
For $\Psi = \sum_a\alpha_a(K)c\beta_a(K) Bc\,\gamma_a(K)$ in the ordinary notation, we have $F(K_1,K_2,K_3) = \sum_a \alpha_a(K_1)\linebreak \beta_a(K_2)\gamma_a(K_3)$.
}
Using this, $\itr{\zeta}\itr{\eta}\Psi$ is calculated as follows:
\def\1{\mathcal I_\eta}
\def\2{\mathcal I_\zeta}
\begin{align}
 	\itr{\zeta}\itr{\eta}\Psi &= \2\l[F_{123}((g/K)_1\identity{12} + \cmt{g/K}{Bc}_{12})(Bc)_{23} + F_{123}c_{12}B_2(g/K)_2\identity{23}\r]\nonumber\\
	&= \2\l[\l(F_{113}(g/K)_1-F_{133}(g/K)_3)(Bc\r)_{13}+(F_{111}(g/K)_1\identity{13})\r] \nonumber\\
	&=-(F_{113}(g/K)_1 - F_{133}(g/K)_3)B_1(h/K)_1\identity{13}\nonumber\\
	&= 0,
\end{align}
where we have used in particular $\itr{\eta}K = \itr{\zeta}K = 0$ in the present case.
Therefore \eqref{eq:W-W} is reduced to
\begin{align}
	\Wilson{C_1}{\xi\p}{\xi} - \Wilson{C_2}{\xi\p}{\xi} &= -\itr{\eta}\itr{\zeta}(\BRST\Psi + \Psi^2)\label{eq:W-W-ex}.
\end{align}
In the restriction $\zeta = (0,h)$ and $\eta = (0,g)$, $\zeta$ and $\eta$ commute each other, $\Lcmt{\zeta}{\eta} = 0$, which is the case in \eqref{eq:w-w} for CS theory.

Next, considering the curve $C$ given by $\xi(s)\,(s\in[0,b])$ connecting $\xi(0) = 0$ and $\xi(b)$,
we will derive the following formula for the Wilson line operated by $\lBRST + \Psi$:\footnote{
In CS theory, the Wilson line \eqref{eq:CS_Wilson} follows the formula $$W_C(x(b),x(a))\l(\overleftarrow{ \td{a}{}{}} + i_{\dot x(a)}A(x(a))\r) = 0 .$$
The largest difference between this formula and \eqref{eq:continue} is that, while the $a$-derivative in the former acts only on the start point, $\lBRST$ in the latter acts on the whole curve $C$.
} 
\begin{align}
	\Wilson{C}{\xi(b)}{0}\l(\lBRST + \Psi(0)\r)&=\Psi(\xi(b))\,\Wilson{C}{\xi(b)}{0}\nonumber\\
	&+i\int_0^b\d s\,\Wilson{C}{\xi(b)}{\xi(s)}\bigl[\itr{\dot\xi(s)}\mathcal F(\xi(s))\bigr]\Wilson{C}{\xi(s)}{0},
	\label{eq:continue}
\end{align}
with $\mathcal F(\xi) := \BRST\Psi(\xi) + \Psi(\xi)^2$.
Here we have introduced the new operator $\lBRST$:
\begin{align}
 	\mathcal A\lBRST := -(-1)^{|\mathcal A|}\BRST \mathcal A,
\end{align}
which has the following properties:
\begin{align}
 	(\mathcal A \lBRST)^\ddag &= -(-1)^{|\mathcal A|}\mathcal A^\ddag\lBRST,\label{eq:leftBRST}\\
 	(\mathcal{AB})\lBRST &= \mathcal A(\mathcal B\lBRST) + (-1)^{|\mathcal B|}(\mathcal A\lBRST)\mathcal B.
\end{align}
The last term in \eqref{eq:continue} vanishes if $\Psi$ is on-shell (then $\Psi(\xi)$ is also on-shell as discussed in section \ref{sec:SFT_in_K}).
The formula \eqref{eq:continue} is expected to be useful for a scenario of degenerate fluctuation modes around a multi-brane solution explained in appendix \ref{app:multi}.

For showing \eqref{eq:continue}, let us discretize the curve $C$ by using the parameter points $s_j = jb/N\,(j = 0,1,\cdots,N)$ to express the Wilson line as a product of Wilson links:
\begin{align}
	\Wilson{C}{\xi(b)}{0} = \lim_{N\to\infty}\wln{\xi(b)}{\xi(s_{N-1})}\cdots \wln{\xi(s_2)}{\xi(s_1)}\wln{\xi(s_1)}{0}.
	\label{eq:discrete}
\end{align}
We start by applying $\lBRST + \Psi$ on $w := \wln{\xi(s_1)}{0} = 1 + i\itr{\xi(s_1)}\Psi$.
Writing $\Psi(0) = \Psi$, $\mathcal F(0) = \mathcal F$, $\xi(s_1) = \xi_1$ and $\Psi(\xi(s_1)) = \Psi_1$, we obtain to $O(b/N)$,
\def\2{\xi_1}
\begin{align}
	 w\bigl(\lBRST + \Psi\bigr) &= \lBRST w - \BRST w + w\Psi= \lBRST w - i\BRST\itr{\2}\Psi + \Psi + i(\itr{\2}\Psi)\Psi\nonumber\\
	&= \lBRST w + (1 + \Lie{\2})\Psi + i\itr{\2}\BRST\Psi + i\itr{\2}(\Psi^2) + i\Psi\itr{\2}\Psi\nonumber\\
	&= \lBRST w + \Psi_1 + i\Psi_1\itr{\2}\Psi+i\itr{\2}\mathcal F\nonumber\\
	&= \bigl(\lBRST + \Psi_1\bigr)w + i\itr{\2}\mathcal F.
\end{align}
Continuing this process to the remaining $N-1$ Wilson links in \eqref{eq:discrete}, we obtain
\begin{align}
 	&\Wilson{C}{\xi(b)}{0}\bigl(\lBRST + \Psi(0)\bigr)=\Psi(\xi(b))\,\Wilson{C}{\xi(b)}{0}\nonumber\\ &\qquad\qquad + i\lim_{N\to\infty}\frac{b}{N}\sum_{j=0}^{N-1}\Wilson{C}{\xi(b)}{\xi(s_{j+1})}
 	\bigl[\itr{\dot\xi(s_j)}\mathcal F(\xi(s_i))\bigr]\Wilson{C}{\xi(s_j)}{0},
\end{align}
which is nothing but \eqref{eq:continue}.

\section{Summary and discussions}\label{sec:sum}
In this paper, we proposed the \KBc interior product $\itr{X}$ and the Lie derivative $\Lie{X}$ specified by a \KBc tangent vector $X$.
By solving the differential equation \eqref{eq:K_sB_sc_s} given by the Lie derivative $\Lie{X}$,
we constructed infinite number of triads $(K(\xi),B(\xi),c(\xi))$ which again satisfies the \KBc algebra.
Using this, we defined the \KBc manifold $\mfd$ consisting of $(K(\xi),B(\xi),c(\xi))$ and having $\xi$ as its coordinate.
Once we get the notion of the manifold, the \KBc interior product, the Lie derivative and the tangent vector are pushed up onto the whole $\mfd$.
On the \KBc manifold, a curve $C$ is parametrized by a real variable $s$ as $\xi(s)$ and the Wilson line along  $C$ can be naturally defined as \eqref{eq:wSFT}.
We found that the Wilson line has the properties \eqref{eq:finite_W_gauge}, \eqref{eq:W-W} and \eqref{eq:continue}.

There remain many questions/problems to be answered.
First, our \KBc manifold is not completely parallel to the ordinary manifold.
One of the largest differences is that quantities carrying ghost number $0$, which seem to correspond to $0$-forms in CS theory, do not vanish by the action of interior products.
This makes the gauge transformation rule of Wilson lines complicated.
As another example, the action of the Lie derivative against vectors \eqref{eq:simple_Lcmt} differs from that in differential geometry.
In addition to this, there is a question about \KBc tangent vectors.
By analogy with the fact that tangent vectors in differential geometry are expressed as $X = X^i\partial_i$, we found the expression of $\hat X$ \eqref{eq:hat} and succeeded in reducing $\Lcmt{X}{Y}$ to $\cmt{\hat X}{\hat Y}$.
Since we adopted $\xi$ as the coordinate of the \KBc manifold, it is strange that the $K$-derivative $\partial$, not the $\xi$-derivative, appears in the first term $X^1K\partial$ of \eqref{eq:hat}.
\modif{One reason for this would be that we started by introducing \KBc interior product, not by defining \KBc tangent vector.
	To find out a way to construct \KBc manifold from the general principle of the manifold is also an important subject.}

As the second problem, the correspondence between SFT and CS theory is incomplete.
Although we introduced the \KBc manifold $\mfd$ and regarded that it corresponds to $M_3$ in CS theory, we have not defined ``integration over $\mfd$."
In fact, the integration $\int$ in the SFT action \eqref{eq:witten} already corresponds to the integration over $M_3$ in CS theory, as given in \eqref{eq:correspondence}.

Thirdly, we comment on the Wilson loop in SFT.
In CS theory, we can construct a gauge-invariant quantity, the Wilson
loop, by considering the Wilson line along a closed loop $C$ and
taking the trace; $\mathrm{Tr}W_C(x(b),x(a))$ with $x(b)=x(a)$.
In SFT, one might be tempted to consider a similar quantity
$\int\Wilson{C}{\xi(b)}{\xi(a)}$ consisting of a Wilson line
for a closed curve $C$ with $\xi(b)=\xi(a)$ and the integration $\int$
giving the SFT action \eqref{eq:witten}.
Indeed, under the gauge transformation, $V(\xi(b))$ and $V(\xi(a))^{-1}$ in \eqref{eq:finite_W_gauge}
cancel each other due to $\int$. However, the extra term in the
exponent in \eqref{eq:finite_W_gauge} persists.
Even worse, since our Wilson line carries no ghost number,
$\int\Wilson{C}{\xi(b)}{\xi(a)}$ vanishes identically; we need an
insertion of $\ghost=3$ quantity to make this non-trivial.
Construction of gauge invariant quantities in SFT\footnote{
See \cite{Hashimoto:2001sm} and \cite{Gaiotto:2001ji} for earlier attempts.}
from the Wilson line by circumventing these difficulties would be an
interesting problem.

Finally, the tools we have found in this paper are restricted to the \KBc subsector of SFT.
However, we expect that they could be generalized to the whole SFT.

\appendix
\section{The determination of $\itr{X}$}\label{app:itr}
Here we determine the \KBc interior product $\itr{X}$ which carries ghost number $-1$ and satisfies the four properties 1.-- 4. mentioned at the beginning of section \ref{sec:itr_Lie_vec}.
Since, at this stage, we do not know that the interior product is specified by a \KBc tangent vector $X$, we write the interior product as $\iitr$ without $X$.

The property 1, the anti-Leibniz rule \eqref{eq:Leibniz}, is just the definition of the operation of  $\iitr$.
Assuming that the actions of $\iitr$ on $K$, $B$ and $c$ is again represented by $K$, $B$ and $c$, they are generically expressed as
\begin{align}
	\iitr K = iBh,\qquad\iitr B = 0,\qquad (\iitr c)_{12} = f_1\identity{12} + ig_{12}[B,c]_{12},
	\label{eq:generic}
\end{align}
where $h = h(K)$, $f=f(K)$ and $g_{12} = g(K_1,K_2)$ are arbitrary complex functions of $K$, and especially $g$
has two variables. For the symbol $(~)_{12}$, see section 2.1 of \cite{Hata_symbol}.\footnote{
For $\iitr c = f(K) + i\sum_{a} L_a(K)\cmt{B}{c}R_a(K)$ in the ordinary notation, we have $g(K_1,K_2) = \sum_a L_a(K_1) \linebreak R_a(K_2)$.
}
Demanding the property 2 for $K$,
\begin{align}
	\iitr K = -(\iitr K)^\ddag =  iBh^\ast(K),
\end{align}
 we find that $h$ is a real function of $K$.
The property 2 for $B$ is trivially satisfied, and that for $c$ gives $f^\ast = f$ and $g_{21}^\ast = g_{12}$, because
\begin{align}
	(\iitr c)_{12} = (\iitr c)^\ddag_{12} = f_1^\ast\identity{12} + i g_{21}^\ast[B,c]_{12}\,.
\end{align} 
Once the property 2 is satisfied for $K$, $B$ and $c$, one can verify that it is also satisfied for any generic product of $K$, $B$ and $c$ due to the anti-Leibniz rule \eqref{eq:Leibniz}.

Next, we examine the property 3. 
The nilpotencies $\iitr^2 K = 0$ and $\iitr^2 B = 0$ are automatically satisfied by \eqref{eq:generic}.
For the condition $\iitr^2 c = 0$, from
\begin{align}
	(\iitr^2 c)_{12} &= \iitr(f_1\identity{12} + ig_{12}[B,c]_{12} )\nonumber\\
	&= iB_1\identity{12}\l(2h_1\textrm{Im}(\l.\partial_1 g_{12})\r|_{2\to 1} 
		+ (\partial_1f_1)h_1 - 2g_{11}f_1\r),\label{eq:nil_c}
\end{align}
obtained by using \eqref{eq:commutators}, we get the following condition:
\begin{align}
	2h_1\textrm{Im}\l.(\partial_1 g_{12})\r|_{2\to 1} 
		+ (\partial_1f_1)h_1 - 2g_{11}f_1 = 0\,.
\end{align}
In \eqref{eq:nil_c}, $\partial_1$ denotes $\partial/\partial K_1$, $(~)_{2\to 1}$ denotes the replacement $K_2\to K_1$, and we have used the relation
$\partial_2 g_{12}\identity{12} = \partial _2 g_{21}^\ast\identity{12} = \partial_1 g_{12}^\ast\identity{12}\,.$

Finally, let us consider the property 4.
The conditions $\iitr([K,B]) = 0$ and $\iitr B^2 = 0$ are trivially satisfied.
For the remaining two conditions, using
\begin{align}
	&\l(\iitr\acmt{B}{c}\r)_{12} = 2ig_{11}B_1\identity{11},\\
	&(\iitr c^2)_{13} = \l(f_1-f_3 + i(g_{11} - g_{33} - 2g_{13}) \r)c_{13} + 2i\l(g_{12} + g_{23}\r)c_{12}(cB)_{23},
\end{align}
we obtain the following three conditions:
\begin{align}
	g_{11} &=0,\qquad
	 f_1-f_2 + i(g_{11} - g_{22} - 2g_{12}) = 0,\qquad
	 g_{12} + g_{23} = \mbox{$K_2$-indep}.
\end{align}

The following conditions are rearrangements of those obtained above:
\begin{align}
	&h^\ast	= h,\quad f^\ast = f,\quad g_{21}^\ast = g_{12}, \quad g_{11} = 0\,,\\
	&h_1\bigl(2\,\textrm{Im}\l.(\partial_1 g_{12})\r|_{2\to 1} + \partial_1f_1\bigr) = 0\,,\label{eq:redundant_cond}\\
	&g_{12} + g_{23} = \mbox{$K_2$-indep}\,,\label{eq:partial_2}\\
	&f_1 - f_2 - 2ig_{12}  = 0\,.\label{eq:ftog}
\end{align}
One can easily verify that the independent conditions are only $h^\ast =h$, $f^\ast = f$ and \eqref{eq:ftog}; other conditions follow from the three.
Since we have 
\begin{align}
 	(\iitr c)_{12} = f_1\identity{12} + \frac{1}{2}(f_1-f_2)\cmt{B}{c}_{12} = f_1\identity{12} + \frac{1}{2}\lcmt{f}{\cmt{B}{c}}_{12},
\end{align}
by using \eqref{eq:ftog}, we obtain \eqref{eq:def_interior} by the identification $X = (X^1,X^2) = (h,Kf)$.

\section{Relation between interior products at different points}\label{app:relation}
In this appendix, we derive the relation between interior products (and Lie derivatives) at 1) infinitesimally separated points $\xi$ and $\xi + \delta\xi$, and 2) the origin and $\xi$.
Here $\delta\xi$ is a constant vector.
\def\1{{\delta\xi}}

For this purpose, we define a map $\mphi{\alpha} : \mfd\to\mfd$ by
\begin{align}
	\mphi{\alpha}(\xi) = \xi - \alpha(\xi),
\end{align}
with $\alpha$ being a \modif{vector field}.
An induced map $\dm{\alpha}$ maps the \modif{vector field} $X$ to another \modif{vector field} $\dm{\alpha} X$, which is defined by the following relation:
\begin{align}
	(\dm{\alpha} X)(\xi) := X(\mphi{\alpha}(\xi)) = X(\xi-\alpha(\xi)).
\end{align}
This map $\dm{\alpha}$ is called differential map in the context of differential geometry.

First, we consider the case 1).
Let $\mathcal O(\xi)$ be a generic product of $K(\xi)$, $B(\xi)$ and $c(\xi)$ at $\xi \in\mfd$.
The operation of $\ITR{X}{\xi}$ on $\mathcal O(\xi)$ is again represented by $K(\xi)$, $B(\xi)$ and $c(\xi)$, so can be expressed as
\begin{align}
	\ITR{X}{\xi}\mathcal O(\xi) = F_{\mathcal O}(K(\xi),B(\xi),c(\xi);X(\xi)).
\end{align}
Applying $1 + \LIE{\1}{\xi}$ against the both hand sides, we get to $O(\1)$,
\begin{align}
 	(\mathrm{l.h.s.}) &\to \ITR{X}{\xi}\bigl(1 + \LIE{\1}{\xi}\bigr)\mathcal O(\xi) + \cmt{\LIE{\1}{\xi}}{\ITR{X}{\xi}}\mathcal O(\xi) 
 	= \l(\ITR{X}{\xi}+\ITR{\Lcmt{\1}{X}}{\xi}\r)\mathcal O(\xi+\1)\label{eq:lhs}
\end{align}
and
\begin{align}
 	(\mathrm{r.h.s.}) &\to F_{\mathcal O}(K(\xi + \1),B(\xi+\1),c(\xi+\1);(1+\LIE{\1}{\xi})X(\xi))= \ITR{\widetilde X}{\xi+\1}\mathcal O(\xi+\1).\label{eq:rhs}
\end{align}
Here we have used \eqref{eq:Lie_itr_cmt} for \eqref{eq:lhs} and defined $\widetilde X$ as
\begin{align}
	\widetilde X(\xi) = \dm{\delta\xi}[(1+\LIE{\1}{\xi})X(\xi)] = X(\xi-\1) + \LIE{\1}{\xi-\1}X(\xi-\1).
\end{align}
This implies the relation
\begin{align}
	\ITR{X}{\xi} + \ITR{\Lcmt{\1}{X}}{\xi} = \ITR{\widetilde X}{\xi+\1}.\label{eq:pre_rewritten}
\end{align}
Then using \eqref{eq:L-L} and the following relation which is valid to $O(\1)$,
\begin{align}
	X = \dm{-\delta\xi}[( 1-\LIE{\1}{\xi})\widetilde X],
\end{align}
and making the replacement $\widetilde X \to X$, \eqref{eq:pre_rewritten} is rewritten as
\begin{align}
	\ITR{X}{\xi+\1} = \ITR{\dm{-\delta\xi}(X - \LIE{X}{\xi} \delta\xi)}{\xi}~.\label{eq:itr_to_itr_all}
\end{align}
The formula \eqref{eq:itr_to_itr_all} with a constant 
vector $X(\xi) = f$,
\begin{align}
	\ITR{f}{\xi+\1} = \ITR{f - \LIE{f}{\xi} \delta\xi}{\xi}~,\label{eq:itr_to_itr} 
\end{align}
 is used to show a property of the Wilson line in subsection \ref{subsec:properties}.
We can show that the same formula holds for the Lie derivative:
\begin{align}
	\LIE{X}{\xi+\1} = \LIE{\dm{-\delta\xi}(X - \LIE{X}{\xi} \delta\xi)}{\xi}~.\label{eq:lie_to_lie}
\end{align}

Next, in order to find the relation for the case 2), we start with applying $\ITR{X}{0}$ on $K(\xi)$, $B(\xi)$ and $c(\xi)$:
\begin{align}
	\ITR{X}{0}K(\xi) &= iB(\xi)X^1(0)(1 + \partial\xi^1),\qquad
	\ITR{X}{0}B(\xi) = 0,\nonumber\\
	\ITR{X}{0}c(\xi) &= \frac{X^2(0) + X^1(0)K\partial\xi^2}{K(\xi)}
	+\lcmt{\frac{X^2(0) + X^1(0)K\partial\xi^2}{K(\xi)}}{B(\xi)c(\xi)}.
\end{align}
Note that these expressions are of the form of \eqref{eq:xi_itr}.
By defining
\begin{align}
	\widetilde X\p(\xi) := (\dm{\xi}X)(\xi) + (\dm{\xi} X)^1(\xi)K\partial\xi = X(0) + X^1(0)K\partial\xi,
\end{align}
the relation
\begin{align}
	\ITR{X}{0} = \ITR{\widetilde X\p}{\xi} = \ITR{\dm{\xi}X + (\dm{\xi}X)^1K\partial\xi}{\xi}~,\label{eq:origin-xi}
\end{align}
holds for $K(\xi)$, $B(\xi)$ and $c(\xi)$.
Using that each of the three expressions of \eqref{eq:origin-xi} follow the anti-Leibniz rule \eqref{eq:Leibniz}, this relation \eqref{eq:origin-xi} holds in the \KBc subsector.
The same relation as \eqref{eq:origin-xi} holds for the Lie derivative:
\begin{align}
	\LIE{X}{0} = \LIE{\dm{\xi}X + (\dm{\xi}X)^1K\partial\xi}{\xi}\,.	
\end{align}

\section{A mechanism of emergence of degenerate fluctuation modes using the SFT Wilson line}\label{app:multi}
In this appendix, as a possible application of our Wilson line in SFT, we present a scenario of the emergence of degenerate fluctuation modes around a multi-brane solution within the \KBc subsector.
See \cite{Erler:2014eqa} for another approach.

\subsection{SFT with Chan-Paton factors}
First, let us consider SFT with Chan-Paton factors, where the string field has indices; $\Psi^{ab}$ ($a,b = 1,\cdots,N$).
This SFT describes the theory of $N$ D25-branes, and each string state has $N^2$ degeneracies.
Using vertex operators $\calO_F(k)$ for each string state $F$ with momentum $k_\mu$ (an example is $\calO_\textrm{tachyon}(k)=e^{-K/2}\,c\,e^{ik\cdot X}e^{-K/2}$), $\Psi^{ab}$ is expanded as
\begin{equation}
\Psi^{ab}=\int\!\frac{d^{26}k}{(2\pi)^{26}}\sum_F\calO_F(k)\varphi_F^{ab}(k),\label{eq:Psi^ab}
\end{equation}
where $\varphi_F^{ab}(k)$ is the associated component field.
The present string field is subject to the reality condition $\bigl(\Psi^{ab}\bigr)^\ddag=\Psi^{ba}
$.
Taking the vertex operator satisfying the condition $\calO_F(k)^\ddag=\calO_F(-k)$, the component field has to satisfy the reality condition 
\begin{equation}
\varphi_F^{ab}(k)^\ddag=\varphi_F^{ba}(-k).\label{eq:phi_reality}
\end{equation}
Then the SFT action with trace over Chan-Paton factors reads 
\begin{align}
S
&=\int\!\tr\left(\frac12\Psi\,\QB\Psi+\frac13\Psi^3\right)\nonumber
\\&=\frac12\int_{k,k'}
\left(\int\!\calO_F(k)\,\QB\calO_{F'}(k')\right)\varphi_F^{ab}(k)\varphi_{F'}^{ba}(k')
\nn\\
&\qquad
+\frac13\int_{k,k',k''}
\left(\int\!\calO_F(k)\calO_{F'}(k')\calO_{F''}(k'')\right)\varphi_F^{ab}(k)\varphi_{F'}^{bc}(k')\varphi_{F''}^{ca}(k''),
\label{eq:S_CP}
\end{align}
where we have omitted $\sum_F$, used the abbreviation $\int_k = \int \d^{26}k/(2\pi)^{26}$, and put $g^2 = 1$.
In the last two terms of \eqref{eq:S_CP}, we have omitted the sign factors\modif{, which arise from the change of the ordering of $\varphi_F^{ab}$\,'s if we include the ghost fields in \eqref{eq:Psi^ab}}
(the sign factors are the same as those in \eqref{eq:with_CP}, which are also omitted there).

\subsection{SFT around a multi-brane solution}
Our problem is whether we can reproduce the action \eqref{eq:S_CP} for the fluctuation $\Delta\Psi$ around a possible $N$ brane classical solution $\Psi_0$, $\Psi=\Psiz+\DPsi$, in SFT \eqref{eq:witten} without Chan-Paton factors.
The action of the fluctuation $\Delta\Psi$ is given by
\begin{align}
\calS
=\int\!\left(\frac12\DPsi\,Q_{\Psiz}\DPsi+\frac13\DPsi^3\right),\label{eq:calS}
\end{align}
where $Q_{\Psi_0}$ the BRST operator around $\Psi_0$ defined by \eqref{eq:BRST_around}.
Here we assume that $\Delta\Psi$ is expanded in terms of $V_a^\ddag\calO_F(k)V_b$ with some $V_a$ $(a = 1,\cdots,N)$ carrying $\ghost[V_a] = 0$ and the associated component field $\varphi_F^{ab}(k)$:
\begin{align}
	\DPsi=\int_k\sum_{a,b}V_a^\ddag\calO_F(k)V_b\,\varphi_F^{ab}(k).\label{eq:DPsi}
\end{align}
The reality condition $\DPsi^\ddag = \DPsi$ implies again \eqref{eq:phi_reality}.
Let us substitute \eqref{eq:DPsi} into \eqref{eq:calS} to examine whether we can reproduce \eqref{eq:S_CP}.
First, from 
\begin{align}
Q_{\Psiz}\!\left(V_a^\ddag\calO_F(k)V_b\right)
&=V_a^\ddag\left(\QB\calO_F(k)\right)V_b
+\left[\bigl(\QB+\Psiz\bigr)V_a^\ddag\right]\calO_F(k)V_b\nonumber\\
&\qquad\qquad -(-1)^{|\mathcal O_F(k) |}V_a^\ddag\calO_F(k)\bigl[V_b\bigl(\rQB+\Psiz\bigr)\bigr],
\end{align}
we find that $V_a$ should satisfy
\begin{align}
	V_a\bigl(\rQB+\Psiz\bigr)=0,\label{eq:first_cond}
\end{align}
which is equivalent to $\bigl(\QB+\Psiz\bigr)V_a^\ddag = 0$ by using \eqref{eq:leftBRST}.
Assuming that $V_a$ satisfies \eqref{eq:first_cond}, we obtain
\begin{align}
\calS
&=\frac12\int_{k,k'}
\left(\int V_a^\ddag\calO_F(k)V_bV_{a'}^\ddag
\bigl(\QB\calO_{F'}(k')\bigr)V_{b'}\right)
\varphi^{ab}(k)\varphi^{a'b'}(k')
\nn\\
&\quad
+\frac13\int_{k,k',k''}\left(\int\!
V_a^\ddag\calO_F(k)V_b\,V_{a'}^\ddag\calO_{F'}(k')V_{b'}
\,V_{a''}^\ddag\calO_{F''}(k'')V_{b''}\right)
\varphi_F^{ab}(k)\varphi_{F'}^{a'b'}(k')
\varphi_{F''}^{a''b''}(k'').
\label{eq:with_CP}
\end{align}
This action is reduced to \eqref{eq:S_CP} by imposing another condition on $V_a$:
\begin{align}	
	V_a V_b^\ddag=\delta_{a,b}\mathbb I\qquad (a,b = 1,\cdots,N).\label{eq:second_cond}
\end{align}

Therefore, the problem is to construct $N$ $V_a$\,'s  satisfying \eqref{eq:first_cond} and \eqref{eq:second_cond}.
First, let us consider \eqref{eq:first_cond}.\footnote{
For a classical solution of pure-gauge type, $\Psi_0 =U\QB U^{-1}$, $V_a = U^{-1}$ is a solution to \eqref{eq:first_cond} since we have $V_a\bigl(\rQB+\Psiz\bigr)=(V_aU)\rQB U^{-1}$.}
The formula \eqref{eq:continue} for the SFT Wilson line suggests us that it could be a candidate for $V_a$ satisfying \eqref{eq:first_cond}.
In fact, the last term of \eqref{eq:continue} with $\Psi = \Psi_0$ vanishes since $\mathcal F(\xi(s)) = 0$.
As for the first term on the r.h.s.\ of \eqref{eq:continue}, $\Psi_0(\xi(b))\Wilson{C}{\xi(b)}{0}$, it could be possible that $\Psi_0(\xi(b))$ goes to zero by taking $\xi(b)$ to the ``infinity" on the \KBc manifold.
For example, let us consider the 2-brane solution \cite{Murata:2011ex, Hata:2011ke,Murata:2011ep,Hata_symbol} given by
\begin{align}
	\Psi_\textrm{2-brane} =-  \frac{1}{\sqrt{K}}\,c\,\frac{K^2}{1 + K}\,Bc\,\frac{1}{\sqrt{K}}.
\end{align}
Then taking $\xi^1\to\infty$, we find that $\Psi_\textrm{2-brane}(\xi)\to 0$.
This is because, from \eqref{eq:KBc_xi}, $\Psi_\textrm{2-brane}(\xi)$ is given as follows:
\begin{align}
	\Psi_\textrm{2-brane}(\xi) &= -\frac{e^{-i\xi^2}}{\sqrt{K(\xi)}}\,c\,\frac{e^{-\xi^1}K(\xi)^2}{1+K(\xi)}\,Bc\,\frac{e^{i\xi^2}}{\sqrt{K(\xi)}} \sim O(e^{-\xi^1}).
\end{align}

Even if  we adopt as $V_a$ the Wilson line extending to the infinity and satisfying \eqref{eq:first_cond}, there still remains a problem; whether there exist $N$ curves $C_a$ satisfying the orthonormality condition \eqref{eq:second_cond}.
Note that $V_aV_b^\ddag$ is a Wilson line of the curve which starts at the infinity, goes along $C_b$ in the reverse direction to reach the origin, and then returns to the infinity along $C_a$.
Therefore, the normalization condition $V_aV_a^\ddag = \mathbb I$ is automatically satisfied.
For establishing the orthogonality, $V_aV_b^\ddag = 0$ for $a\neq b$, we need a deeper understanding of the \KBc manifold.

\bibliographystyle{jhep} 
\bibliography{}
\end{document}